Banner appropriate to article type will appear here in typeset article

# Plunging Breakers - Part 2. Droplet Generation

**M. A. Erinin**[1,*], **C. Liu** [1], **S. D. Wang**[1], **X. Liu** [1] **& J. H. Duncan** [1]†,

[1]Department of Mechanical Engineering, University of Maryland, College Park, MD 20770, USA

[*]present address: Department of Mechanical and Aerospace Engineering, Princeton University, Princeton, NJ 08544, USA

The positions, diameters ($d \geqslant 100$ µm), times and velocities of droplets generated by three plunging breaking waves are measured as the droplets move up across a measurement plane located 1.2 cm above the crests during breaking. The three breakers are created by dispersively focused wave packets that differ primarily by small differences in the overall amplitude of the wave maker motion used to generate them. The breakers are designated qualitatively by their intensities: weak, moderate and strong. The droplets are measured with two in-line cinematic holographic systems operating at 650 holograms per second with measurement volumes that span the width of the tank.

It is found that there are four major mechanisms for droplet production: closure of the indentation between the top surface of the plunging jet and the splash that it creates (in a spatio-temporal region labeled Region I-A), the bursting of large bubbles that were entrapped under the plunging jet at impact (Region I-B), splashing and bubble bursting in the turbulent zone on the front face of the wave (also in Region I-B) and the bursting of small bubbles that reach the water surface at the crest of the nonbreaking wave following the breaker (Region II). The droplet diameter distributions for the entire droplet set for each breaker are fitted with power law functions in separate small- and large-diameter regions. The droplet diameter where these power law functions cross increases monotonically from 820 µm to 1480 µm from the weak to the strong breaker, respectively. Similar power law behavior is found in Regions I-A. The average droplet 2D speed (computed from the streamwise and vertical velocity components) in Region I-A over the three breakers is 1.02 m/s and the average 2D velocity direction is about 8.9° downstream from vertical (downstream is defined herein as the direction of wave travel). In Region I-B the average 2D speed and velocity direction are 1.21 m/s and 33.9°, respectively, while they are 0.74 m/s and 15.9°, respectively, in Region II. The mechanisms that create these differing droplet characteristics are discussed with the aid of the profile measurements from Part 1 of this two-part paper. Results of a simple calculation of the droplet motions are presented that, together with the droplet data, indicate the strong influence of the breaker-induced air motion on the droplet trajectories.

**Abstract must not spill onto p.2**

# 1. Introduction

As mentioned in the introduction to Part 1, sea spray is a natural phenomenon with wide ranging implications in the transfer of mass, momentum and energy between the ocean and the atmosphere, see for example Andreas (1992), Melville (1996) and Andreas (2002). These spray-droplet-augmented transfer processes can have a dramatic impact on the weather, climate (Andreas & Emanuel (2001)) and chemical reactions that take place near the water surface, see for example Galbally *et al.* (2000) and Prather *et al.* (2013). It is widely accepted that breaking waves play a key role in the production of sea spray droplets. In breaking events (spilling or plunging), droplets are thought to be produced by splashing, wind shear, and through the bursting of breaker-entrained air bubbles as they return to the water surface. The generation mechanisms of sea spray droplets have been widely studied, and Veron (2015) has given a thorough review of the state of knowledge of sea spray droplet generation and behavior.

A number of studies have focused on the vertical distributions of droplet characteristics in wind wave systems in the field and the laboratory. In the field, Wu *et al.* (1984) conducted measurements of the droplet size spectrum in the Delaware Bay for droplet diameters $d > 50$ µm and found that the normalized droplet diameter distribution consisted of two linear regions when plotted on a log-log plot. de Leeuw (1986) measured droplets ($10 \leqslant d \leqslant 100$ µm) in the North Atlantic at wind speeds up to 11 m/s and at heights up to 11 m above the local sea level. From the data, the effect of wind speed on the function describing the variation of droplet concentration and diameter with vertical height was explored. Smith *et al.* (1993) measured droplets ($2 \leqslant d \leqslant 50$ µm) off the west coast of Scotland for wind speeds up to 30 m/s and quantified the relationship between droplet production rate and wind speed. In the laboratory, Wu (1979) studied the problem in a wind wave tank where he found the droplet diameter distribution curve plotted in log-log coordinates drops off rapidly for $d > 200$ µm, in qualitative agreement with his above referenced field experiments in 1984. Koga (1981) studied the movement and production mechanism for droplets generated by wind waves in light wind conditions. Anguelova *et al.* (1999) observed sprays of spume droplets tearing off wave crests and measured droplets with diameters in the range $1.3 \leqslant d \leqslant 10.5$ mm. Veron *et al.* (2012) studied spray generation at high wind speeds (31.3 to 47.1 m/s) and found that the numbers of large droplets exceeded theoretical predictions. More recently, Erinin *et al.* (2022) measured droplet speed and acceleration statistics of spray generated at wind speeds up to 12 m/s in a wind-wave field in a laboratory tank and reported on droplet speed and acceleration probability density functions. The authors found droplets with speeds greater than the measured wind speed. In related experiments, Ramirez de la Torre *et al.* (2022) reported measurements of droplet statistics in laboratory experiments with and without wind as focused wave packets approached a shoal.

A model accounting for the droplet production rate with wind speed was first proposed by Monahan *et al.* (1986) and refinements and additional models were proposed by de Leeuw (1986), Andreas (1992), Wu (1990) and Andreas *et al.* (1995). However, in a review by Andreas (1998) it is pointed out that estimates of the production rate of droplets can vary over six orders of magnitude.

In the review article, Veron (2015) postulates that there are two primary generation mechanisms for droplet generation in the ocean. The first is due to bubbles, initially entrained by the breaking processes, rising to the free surface and popping. The popping bubble generates two different types of drops, film and jet drops. Film droplets are generated when the bubble film fragments creating many droplets with diameters reported to range from 20 nm to 200 µm in experiments, see Lhuissier & Villermaux (2012). In these experiments, jet droplets were observed to form as a result of the violent bubble cavity collapse which ejects up to six droplets ranging in size from 2 to 200 µm. Results of similar experiments were reported by Wu (2002) and others. More recently, revised scaling arguments for the size of droplets generated by bursting bubbles have been presented by Gañán Calvo (2017). Deike *et al.* (2018) studied the conditions, including the speed of upward traveling jet, under which a jet droplet is observed by using experimental and numerical results. They found that the jet ultimately controls the velocity of the resulting droplets. It is important to emphasize that these studies primarily involve single bubble bursting events in calm water, while in a breaking wave bubble bursting occurs in a dense field of bubbles that rise to the surface in an unsteady turbulent flow.

The second primary mechanism for droplet generation occurs when the wind speed above the breaking wave is sufficiently high to tear off water from the crest of the wave. These "spume droplets" are thought to be largest droplets, $d \gtrsim 1.0$ mm, measured in the above-mentioned wind wave systems in the laboratory and the field. Tang *et al.* (2017a) developed a direct numerical simulation (DNS) scheme to study the generation and transportation of spume droplets by wind blowing over breaking waves. They found that droplets are generated near and/or behind the wave crest, depending on the wave age. To date, simulations are only able to resolve relatively large droplets, and often the droplets are represented by points in the computations.

Additional mechanisms of droplet generation in breaking waves have also been addressed. Veron (2015) identified the droplets that may be produced by the plunging jet impacting on the free surface, although it is believed that this mechanism is not as efficient at producing droplets. Also, Lubin *et al.* (2019) discussed the instabilities which may be responsible for air-entrainment and droplet generation in breaking waves.

Droplet production by breaking waves generated without wind in CFD and laboratory wavetanks has also been explored. In the CFD investigations, DNS calculations are performed with the domain covering one streamwise wavelength ($\lambda$) of a periodic uniform wavetrain and the initial wave surface and flow field in the water are taken from 3rd order Stokes wave theory. The initial wave steepness is chosen to be excessive for 3rd order theory and the wave evolves to a plunging breaker. The dynamics of these breakers has been explored by a number of authors, see for example the 2D studies by Chen *et al.* (1999) and Iafrati (2009). Droplet generation in breaking Stokes wavetrains is explored using 3D DNS in the work of Wang *et al.* (2016) and Mostert *et al.* (2022). These studies are important numerical counterparts to the experiments presented herein. In Wang *et al.* (2016), energy dissipation, air entrainment and droplet generation are explored. The authors identify droplet production mechanisms similar to those discussed herein and provide droplet diameter distributions that are fitted well with a power law function. From the dimensionless parameters given in the paper, the initial amplitude is 2.4 cm (given the wave slope $\epsilon = 0.55$), the gravity wavelength in the calculations is 27.3 cm, and the minimum computational grid spacing is 65 µm. For a droplet resolved with five grid points across its diameter, the minimum resolved droplet diameter would be 0.26 mm. In Mostert *et al.* (2022), the discussion stresses the computed droplet diameter and velocity distributions as well as the temporal histories of the droplet generation process. Scalings based on measurements of the wave profile at the moment of jet impact are explored. From the dimensionless parameters given in the paper, the initial wave slope is 0.63, the gravity wavelengths for the two high-Reynolds number droplet diameter PDFs given in figure 15 of the paper are 38.3 cm and 54.2 cm, and the grid resolutions are 0.188 mm and 0.265 mm, respectively. Thus, with a droplet spanning 5 grid points across its diameter, droplets with diameters as small as 0.94 and 1.33 mm, respectively, are resolved, though results for smaller droplets are presented.

In the present two-part paper, the profile histories (Part 1) and droplet generation (Part 2) in three plunging breakers are studied. The breakers are generated mechanically by wave maker motions that differ primarily by the overall amplitude of their height versus time profiles. In Part 1, the profile histories are recorded for an ensemble of 10 breaking events for each wave and the data is used to create temporally evolving ensemble average and fluctuating (using two measures of profile standard deviation) profile histories of each the three breakers. In these three waves, the size of the plunging jet and the region under it grows with wave maker amplitude as does the apparent intensity and the measured rms fluctuations of the breaking region after jet impact. For this reason, these three waves are called the weak, moderate and strong breakers, in reference to a quality referred to as the breaking intensity. In Part 2 (the present paper), cinematic holography-based droplet measurements recorded as the droplets move up across a measurement plane locate just above the breaking crests are presented and discussed. The droplet measurements consist of their numbers, positions, times, diameters and two components of their velocities. These droplet data are interpreted with the aid of the profile data from Part 1 to determine the local breaking processes that generate droplets in each wave and the characteristics of the droplets associated with each breaking process. To the authors' knowledge, this is the first study (including the preliminary results for the weak breaker in Erinin *et al.* (2019)) to explore the connections between droplet characteristics and breaking mechanisms, as identified and measured in wave profile measurements from an ensemble of breaking events.

In the following, the experimental details of the cinematic in-line holography measurements of the droplets are presented first in § 2. This is followed, in subsections § 3.1 to § 3.5, by descriptions and discussions of the droplet measurement results. Finally, the conclusions of this study are discussed in § 4.

## 2. Experimental Details

The droplet measurements were performed in the same facility and with the same three waves as in the wave profile measurements described in Part 1 of this two-paper sequence. The Part 1 paper includes descriptions of the wave tank, wave maker, wave maker motions, instrument carriage, experimental procedures and the analysis of the profiles of the three waves studied herein. For each of these three waves, the average wave packet frequency was $f_0 = 1.15$ Hz ($T_0 = 1/f_0 = 0.870$ s). In this section, only the droplet measurement techniques are described.

### 2.1. *Droplet Measurements Using In-line Holography*

The droplets generated by the three breaking waves are measured with two synchronized identical cinematic in-line holographic systems, see figure 1 for details. The two systems are attached side by side to the instrument carriage with a horizontal distance of 40.6 cm between their optical axes. The bottom edges of the images are horizontal and located at the same height, 1 cm above the highest height reached by the breaking crest surface for each wave. These wave heights are (from table 2 of Part 1) 107.8, 110.6, and 111.5 mm for the weak, moderate and strong breakers, respectively.

The laser pulses and cameras are synchronized to take holographic images at a rate of 650 pps for a duration of 1.974 s ($2.270T_0$) starting at approximately the time of jet impact in each breaking event. Holographic image sequences are taken at 28 streamwise locations by moving the carriage to 14 fixed positions. The results are interpolated to cover regions between measurement locations where no hologram image sequences are recorded. The droplet measurement locations cover a streamwise region from just before the jet impact site to approximately 1 meter downstream. At each location, 10 experimental runs were performed for a total of 140 individual breaking events for each of the three breakers. At two locations, droplets from an additional 32 breaker realizations were performed in order to determine the number of realizations required for measurement convergence. It was found that 10 realizations created a good balance between the measurement/hologram calculation time and the desire for good data convergence.

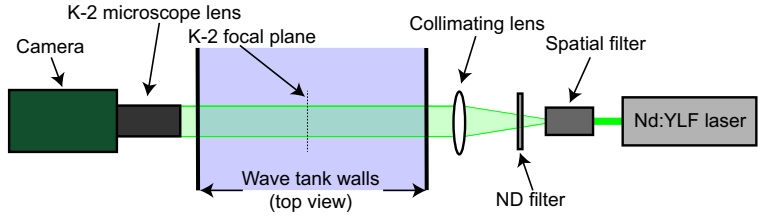

Figure 1: A schematic drawing of one of the two side-by-side channels of the droplet hologram recording system. All components are mounted on the instrument carriage. A light beam is generated by a pulsed Nd:YLF laser. The beam passes through a spatial filter, a neutral density filter and a collimating achromat lens, thus creating a 50-mm diameter nearly uniform horizontal beam that crosses the width of the tank in the direction perpendicular to the tank side walls. The holographic images are captured by a Phantom V640 camera (4 Megapixel, 12 bit images) that is fitted with a K-2 long-distance microscope lens (Infinity Photo-Optical Company). The optical axis of the lens is aligned with the centerline of the laser beam and the lens is focused on the vertical center plane of the tank; the image magnification is 1:1 (10 µm/pixel). Two configurations of the light source were used under different experimental conditions: 1.) a single high-energy Nd:YLF laser (50 mJ/pulse, 150 ns pulse width, Photonics Ind., DM50-527 and combined with two beam splitters used to dump approximately 99.8% of the light energy) was used for droplet measurements for the weak and strong breakers. The final expanded beam was split in two to create light beams for the two cameras; and 2.) two lasers (80 µJ/pulse, 20-30 ns pulse width, CrystalLaser, model QL527-20) were used for the droplet measurements with the moderate breaker, one laser for each channel. Additional details can be found in Erinin (2020).

Each of the holographic images are processed digitally in the manner described below and in detail in Erinin (2020). First, the background image (recorded before the breaking event in each run) is subtracted from the recorded hologram to remove interference patterns from system imperfections, dust and water droplets on the optical surfaces and tank walls. The resulting hologram is digitally reconstructed every 5 mm in the $z$ (cross-tank) direction using the Fresnel-Huygens paraxial approximation (Katz & Sheng 2010) via a GPU-based reconstruction algorithm provided by Professor Joseph Katz from Johns Hopkins University. Droplet image volumes are located in this 3D image space and at each location a 200-by-200-pixel window around the diameter of the droplet is reconstructed digitally every 500 µm in $z$. A method similar to the one outlined in Guildenbecher *et al.* (2013) is used to determine the $z$ plane of best focus of the droplet, which is then taken as the $z$ position of the droplet center. The hologram is then reconstructed at this $z$ and the droplet's diameter and the $x$-$y$ position of its center are measured with a custom method using an inverse hyperbolic tangent function that is fitted to the intensity profile of the droplet, see Erinin (2020). It is estimated that the accuracy of measuring the droplet positions is ±10 µm in the $x$-$y$ plane and ±5 mm in $z$. Because of difficulties in reconstructing and accurately measuring the size of droplets near the image boundaries, where portions of the droplet's diffraction pattern are cut off, only droplets measured when they are at least 200 pixels inside the boundaries of the 2560 x 1600 pixel images are counted in the data set. The horizontal plane containing the lower boundary of this inner rectangle in the images is referred to in the following as the measurement plane. This plane is located 2 mm (200 pixels × 10 µm/pixel) above the bottom edge of image, i.e. 1.2 cm above the highest height reached by the breaking crest surface for each wave. A schematic drawing showing the hologram measurement locations relative to a single breaking wave profile is given in figure 2.

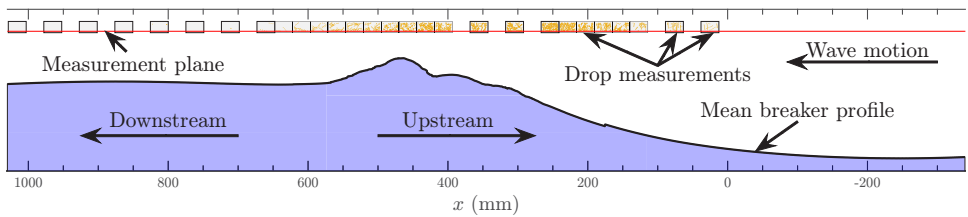

Figure 2: Droplet measurement locations (rectangles) with droplet positions (gold dots inside rectangles) and the ensemble average wave profile (solid black line) at a single instant in time shortly after jet impact. All 28 droplet measurement locations are shown. The solid horizontal red line represents the measurement plane. The relative size between the breaker profile and holographic measurement windows is preserved. The direction of wave propagation is right to left, away from the wave maker (downstream).

In order to calibrate the system and ensure that even the small droplets can be reconstructed and measured accurately across the entire width of the tank, holograms of a custom target are recorded with the target positioned at various locations across the tank width. The calibration target consists of a glass slide with 14 black chrome sputter deposited circles of diameters ($d_c$) ranging from 30 to 3000 µm. A motorized linear traverser is used to accurately place the target at a range of $z$ positions where calibration holograms are recorded. The diameters and $z$ positions of the circles are then measured from the reconstructed holograms. The errors in the hologram-based measurements are then assessed by comparing the known and measured values of $d_c$ and $z$. The maximum droplet diameter measurement error is determined by reconstructing the calibration target at the farthest $z$ distance from the focal plane and is found to be less than 2.5 percent. Thus, each hologram provides the diameters and 3D positions of all of the droplets with diameters $\geqslant 100$ µm within the imaged volume at all positions across the entire width of the tank.

It should be kept in mind that the LIF water surface profile images described in Part 1 are captured over a time interval of approximately 1/650 s in the 1-mm-thick laser light sheet at the streamwise centerplane of the tank while the droplet positions and diameters are captured over the duration of the Nd:YLF laser light pulse, approximately 30 ns, and throughout the entire width of the tank.

### 2.2. *Droplet Tracking and Velocity Determination*

The droplets from the breaking wave are tracked in time as they move through the 3D image space by using a tracking algorithm based on a modified nearest-neighbor algorithm developed for Brownian particle motion by Crocker & Grier (1996), see Erinin (2020) for details. Each trajectory is then fitted with separate second order polynomials for $x_d(t)$, $y_d(t)$ and $z_d(t)$. The polynomials are then used to find the time, $x$ and $z$ positions, and velocity of each droplet as it crosses the measurement plane. Only droplets that are moving up through the measurement plane are included in the data set. Droplets that are moving downward through the measurement plane or into the 3D image space through its top or side surfaces are assumed to have been accounted for moving up across the measurement plane at other measurement or interpolated positions. The upper limit of measurements of the droplet vertical ($y$) and streamwise (horizontal, $x$) velocity components is determine by the smallest image dimension (the 1,600 pixel height), the pixel resolution (10 µm), the image frame rate and the requirement that the droplet be imaged three times in order to be tracked. Considering these constraints, it is estimated that the maximum vertical velocity of a droplet must be below approximately 4 m/s. In practice, very few droplets with vertical speeds greater than 3.5 m/s were found. The measurements of the cross stream position ($z$ coordinate) of the droplet position is inaccurate as noted in §2.1, and this leads to even greater inaccuracies in the $z$-component of the droplet velocity. Thus, only the streamwise ($u$, positive in the direction of wave travel) and vertical ($v$, positive up) droplet velocity components are reported and discussed below.

### 2.3. *Humidity and Droplet Evaporation*

Estimates of the droplet diameter decrease due to evaporation during the time of flight from the location of droplet generation at the free surface to the measurement plane were estimated via the theory of Pruppacher & Klett (1978). In order to estimate this change in diameter, the humidity in the wave tank is required. To this end, a humidity sensor (Thorlabs, model TSP01) was placed in the wave tank at the level of the measurement plane for the moderate breaker experiments. The minimum relative humidity was found to be approximately 50%. Using this value in the theory, it was found that the decrease in droplet diameter would be less than 4% after 1 s for a droplet with an initial diameter of $d = 100$ µm and therefore insignificant for the 70-ms maximum flight time estimated from the present wave profile and droplet velocity data, see § 3.5.

### 2.4. *Surface Tension*

The surface tension isotherm of samples of water from the wave tank were measured, as described in Part 1, twice per day during all of the droplet measurement experiments. In all cases, the surface tension remained at the clean-water value of 73.0 dyne/cm through compression of 80%. See Part 1 for further details.

## 3. Results and Discussion

The presentation and discussion of the results is divided into five subsections with the spatio-temporal distributions of the numbers and diameters of the droplets measured over the breaking crests in § 3.1, the ensemble averaged spatio-temporal contour maps of the local number of droplets over the entire measurement field in § 3.2, the droplet diameter distributions in § 3.3, the droplet velocity distributions in § 3.4 and a low-order droplet motion model in § 3.5. To assist with the verbal descriptions in this section, four white light movies are given as supplemental material. Movie 1 includes above and below surface views of the strong breaker and uses stop action and closeup views to identify the various droplet generation mechanisms. Movie 2, Movie 3 and Movie 4 consist of above and below surface views of the strong, moderate and weak breakers, respectively, and provide uninterrupted images sequences of each breaking event.

### 3.1. *Droplet Distribution over the Breaking Crest*

In order to explore the relationship between events in the breaking process and the spatio-temporal distribution of droplet generation over the breaking crests, the position, time and diameter of each droplet measurement are indicated on top of the crest profiles for the weak, moderate and strong breakers in figures 3(*a*), (*b*) and (*c*), respectively. The profiles are ensemble averages over 10 realizations of each breaker that are taken from figure 12 of Part 1 and plotted in the laboratory reference frame. The droplet data is from three breaker realizations at each measurement location. Plotting only a subset of the droplet data was necessary because higher numbers of droplets from additional breaker realizations made the plots difficult to interpret. Further details of the plotting scheme are given in the figure caption.

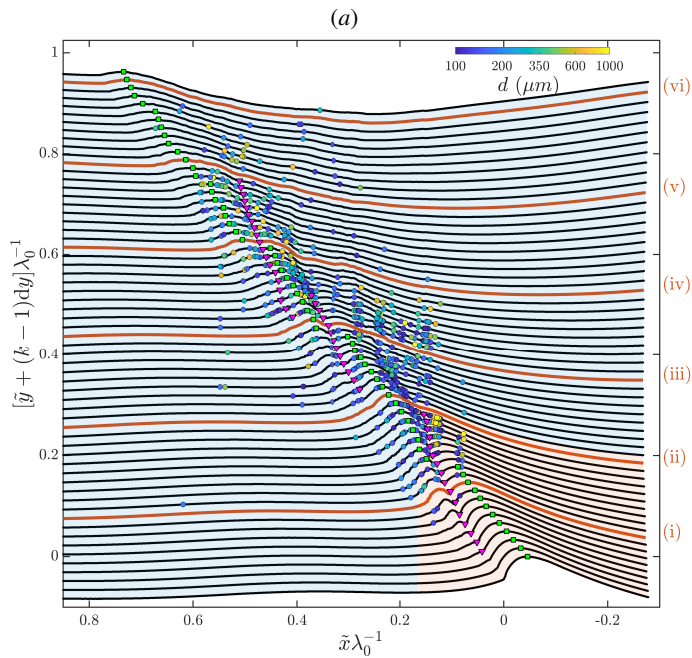

Figure 3: Subplot (*a*). The spatial distributions of droplets plotted on top of the evolution of the ensemble average surface profiles for the weak, moderate and strong breakers are presented in subplots (*a*), (*b*), and (*c*), respectively. The ensemble average surface profiles are the same as those shown in figure 10 of Part 1; however, the profiles shown here are plotted in the laboratory reference frame. A similar version of subplot (*a*) was presented in Erinin *et al.* (2019). The profiles and droplets are plotted relative to the ensemble average position and time of jet impact, $\tilde{x} = 0$ and $\tilde{t} = 0$, respectively. The bottom-most profile in each plot is the crest shape at $\tilde{t} = 0$ and each successive profile (the time between profiles is $\Delta t = 12.3$ ms) is plotted d$y = 20$ mm above the previous. Every tenth profile is colored red and labeled with a lower case Roman numeral. The green squares are located at the highest point on each surface profile and the magenta upside-down triangles are located at the first three indentations, see the caption for figure 10 in Part 1 for more details. The droplets are plotted as filled circles on the profile recorded at the time closest to the time when the given droplet crosses the measurement plane and are located on that profile at the streamwise position of the droplet crossing. The vertical bands with no droplets appear at locations where droplet data was not collected , see figure 2. These bands are sometimes difficult to see in the present subplots due to the crest point and indention markers and an optical illusion created by the wavy surface profiles. The color of the droplets indicates their diameter as given by the logarithmically scaled color bar. Regions with large numbers of droplets with similar diameters take on a generalized color and may correspond to particular processes in the breaking events. The blue and red colored backgrounds between the profiles show the spatio-temporal limits of two droplet-producing regions, Region I-A (in orange) and I-B (in blue).

Since the horizontal droplet measurement plane is located slightly above the wave crest, see § 2.1, the droplet measurement times and positions are not the same as the positions and times when the droplets were generated at the water surface. The motion of a given droplet between the time it is generated and measured is influenced by the initial droplet ejection velocity, the initial vertical distance from the measurement plane, and the interaction between the droplet and the air flow field, which is created by the breaking wave. Thus, the droplets measured in this experiment were generated at times earlier than the measured times and may have been generated upstream or downstream of their plotted location. These issues will be discussed further in § 3.5. Finally, it should be kept in mind that the droplets are measured across the entire width of the tank while the individual wave profiles are measured only at the center plane of the tank. In spite of this limitation of the profile measurements, the ensemble average profile and distributions of standard deviation should be independent of cross-tank position, except very near the tank walls.

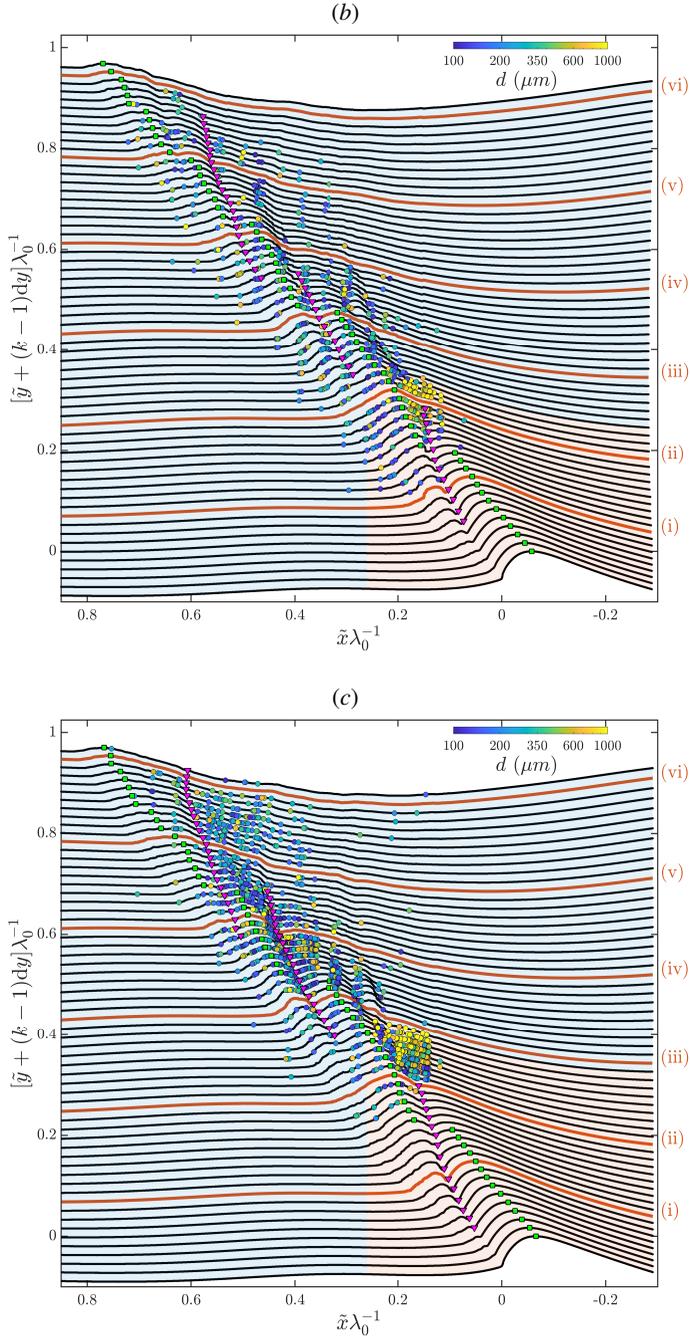

Figure 3: Subplots (*b*) and (*c*). See figure 3 (*a*) for detailed caption.

As can be seen in figures 3(*a*), (*b*) and (*c*), the first droplets are typically measured just before or after the profiles marked (*i*), which corresponds to $\tilde{t}$ = 110.8 ms. Near profile (*ii*), a region begins where many droplets with diameters as large as approximately 1 mm cross the measurement plane. For the strong breaker, this region is sharply defined (measuring approximately 150 mm by 98 ms) and begins at approximately profile (*ii*). The string of inverted magenta triangles marking the first indentation ends on this profile, indicating that the indention is no longer detectable at later times. (As discussed in detail in Part 1, the first indentation forms the boundary between the top surface of the plunging jet and the splash that it creates.) This region of high droplet number is located on the back face of the wave near the crest. As the breaker strength is decreased, the high-droplet-number region becomes more diffuse and begins at earlier times (profiles), before the end of the first indentation. The surface profile normal standard deviation, $n_{sd}$, and the surface profile arc length standard deviation, $s_{sd}$, (both defined in Part 1 and in the caption of figure 6 in the present paper) show only moderate values at the time and position ranges of this intensive droplet flux, see Part 1, figures 12 and 13. As discussed in Part 1, this is the region where the deep crater in the bottom of the first indentation pinches off close to its deepest point and retracts rapidly toward the free surface leaving a small region of bubbles in front and near the bottom of the tube of air entrapped at jet impact, see Movie 1 given as supplemental material.

Figure 4 contains, two sequences of three images, taken from Movie 2 (the strong breaker), that show the moments before, during and after the indentation closure. The images in the top row (*a*, *c*, *e*) and bottom row (*b*, *d*, *f*) are from the camera views take from above and below the water surface, respectively. Images (*a-b*) show the indentation 147 ms after jet impact and just before the crater below the indentation closes. By this time, the crest point (the highest point on a given wave profile) is located on the splash-up generated by the plunging jet impact. In the below-surface image, the view of the roller of air entrained at the moment of jet impact is nearly obscured by the crater that extends downward from the indentation at the surface. Only 4 ms later, images (*c*) and (*d*), the indentation crater has nearly reached full retraction as shown in image (*d*) and leaves behind a small number of air bubbles, which can be more clearly seen in Movie 2, Movie 3 and Movie 4. In images (*e*) and (*f*), taken 182 ms after jet impact, the indentation crater is completely retracted and droplets are being ejected all along the indentation. From the plots in figure 3, it can be seen that the number of droplets in this crater closure region increases dramatically with breaker strength. Also, in view of the fact that the region of high droplet number is small, well defined and close to the crater location, it is likely that the droplets' initial velocities are nearly vertical and that the free surface in this region is close to the measurement plane. Droplet generation during the closure of the indentation is also seen in the numerical simulations of Wang *et al.* (2016) and Mostert *et al.* (2022).

Another region of intense droplet flux through the measurement plane is found over the breaking wave crest between profiles (*iii*) and (*vi*). These droplets are most likely generated by two breaking processes: splashing and bubble popping near the leading edge of the breaking zone on the downstream side of this high droplet flux region and the bursting of the large bubbles on the back face of the wave on the upstream side of the region. These large bursting bubbles on the back face are from the air entrapped under the jet at the moment of jet impact. Both of these droplet ejection processes can be seen clearly in the movies given as supplemental material. Also, both regions associated with these processes contain some of the highest values of $n_{sd}$ and $s_{sd}$, see figures 12 and 13, respectively, in Part 1 and the discussion of figure 6 in § 3.2 of the present paper.

For later analysis of droplet number, diameter and velocity distributions, the droplet generation shown in figure 3 is broken into two regions. The first region, called Region I-A and marked by the orange background, includes the jet impact, formation of the first splash, and the formation and closing of the first indentation. The second region, called I-B and indicated by the blue background, covers the remaining regions of the breaking crest and includes the subsequent sequence of splash impacts and splash ups as well as the emergence and bursting of the large bubbles that were entrapped at the moment of jet impact. The spatial and temporal boundaries of Regions I-A and I-B are given in Table 1.

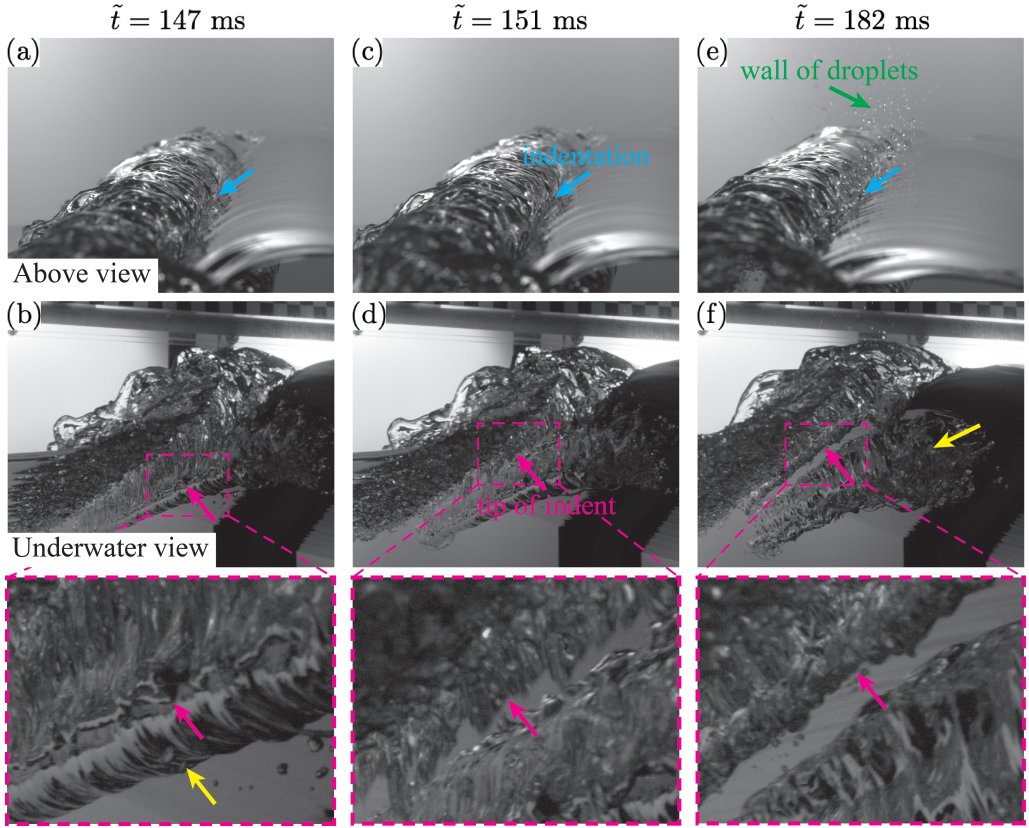

Figure 4: Three white-light image pairs in a time-sequence of the closing of the first indentation with views from above the water surface (images *a*, *c*, and *e*) and below the water surface (images *b*, *d*, and *f*) for the strong breaker. The cyan arrows point to the indentation in each above-surface image and the magenta arrows point to the tip of the indentation crater in the below-surface images. Each pair of images (*a* and *b*), (*c* and *d*), and (*e* and *f*) was captured simultaneously 147 ms, 151 ms and 182 ms after jet impact, respectively, and the first and last image pairs approximately correspond to profiles (*ii*) and (*iii*) in figure 3(*c*). See the text for more details. These images were taken from Movie 2. See also the edited version of this movie, Movie 1. Both movies are given as supplementary material.

### 3.2. *The Distribution of Droplets over the Entire Wave Field*

As discussed above, the position ($\tilde{x}$), time ($\tilde{t}$), diameter ($d \geqslant 100$ µm) and 2D velocity ($\vec{v} = u\hat{i} + v\hat{j}$) of each droplet is measured as it travels upward across the measurement plane. Here we define the droplet number distribution function $N(\tilde{x}, \tilde{t}, d, u, v)$ to be the number of droplets per breaking event per meter of crest length in bins centered on the values of the five independent variables within the ranges $0 \leqslant \tilde{x} \leqslant 1050$ mm, $0 \leqslant \tilde{t} \leqslant 2000$ ms, $d \geqslant 100$ µm, $-3 \leqslant u \leqslant 3$ m/s and $0 \leqslant v \leqslant 3$ m/s. In the following, we present results from various integrations of the distribution function over one or more of these independent variables. For notation, the distribution function is always presented with the independent variables that remain after the integrations. For example, the local number of droplets of all diameters and velocities per breaking event per meter of crest length is written $N(\tilde{x}, \tilde{t})$ and the total number of droplets per breaking event is $N$.

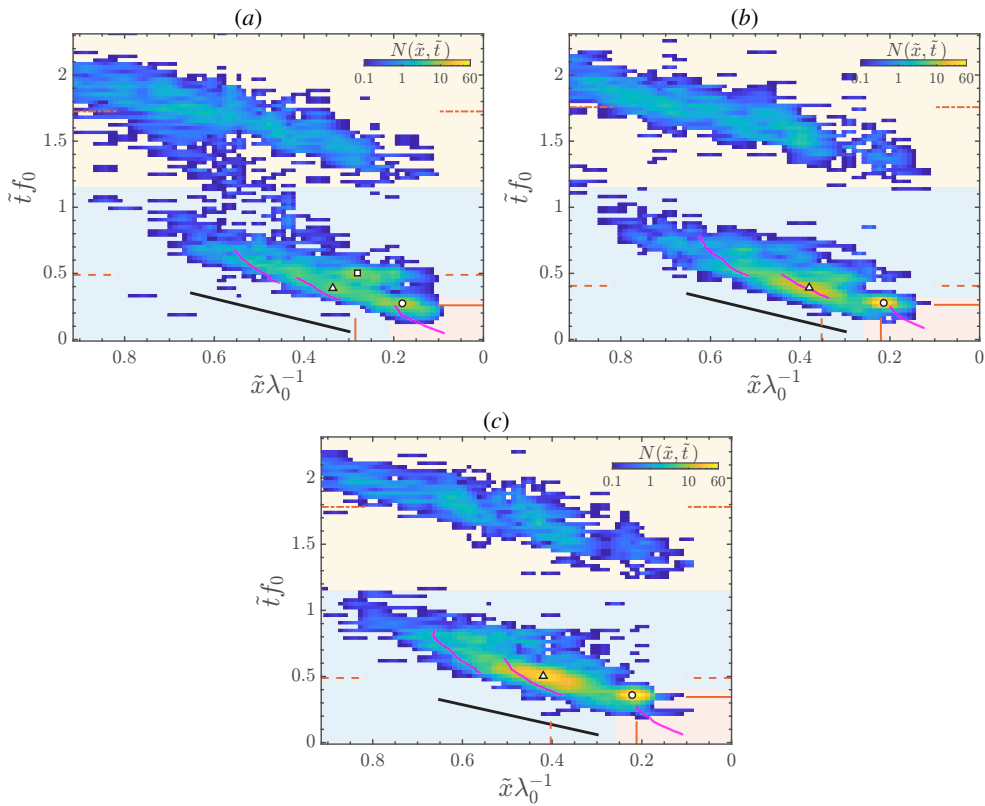

Figure 5: Contour maps of $N(\tilde{x}, \tilde{t})$, the number of droplets moving up across the measurement plane per surface area (m$^2$) per ms per breaking event are shown for the weak, moderate and strong breakers in subplots (*a*), (*b*), and (*c*), respectively. The data is from at least 10 breaker realizations at each droplet measurement location and from interpolation in $\tilde{x}$ intervals where no data was recorded. The contour maps are shown in the laboratory reference frame and cover the full measurement region, $\approx 1050$ mm in streamwise distance and $\approx 2000$ ms in time with a resolution of 13.02 mm by 25 ms. Only droplets with $d \geqslant 100$ μm are counted. Spatio-temporal bins where $N(\tilde{x}, \tilde{t}) \leqslant 0.05$ are colored solid orange, light blue and tan in droplet producing Regions I-A, I-B and II, respectively. The magenta curves mark locations of the three indentations. The horizontal and vertical orange lines indicate the locations of the maxima of $N(\tilde{t})$ and $N(\tilde{x})$, respectively, see figure 7, with the solid, dashed and dotted-dashed lines for the first, second and third local maxima, respectively. The solid black lines are drawn with their slopes corresponding to, $\langle u_{toe} \rangle$, the average speed of the toe shortly after jet impact as computed from the data plotted in figure 10(*a*) of Part 1. For reference, the last wave profile shown in each of the three subplots of figure 3 was recorded at $\tilde{t} = 0.851 f_0^{-1}$. The white filled black circle, black triangle and black square in each plot mark the locations of the local maxima at the end of the first indentation, between the first and second splash ups and the bursting of large air bubbles on the back face of the wave, respectively.

Contour plots of $N(\tilde{x}, \tilde{t})$ are given for the weak, moderate and strong breakers in figures 5(*a*), (*b*) and (*c*), respectively. See the figure caption for exact definitions and details. The most striking features of these droplet number contour plots are the two large spatio-temporal regions of droplet production tilted with a slope close to the speed of the toe of the wave shortly after jet impact, $\langle u_{toe} \rangle$, and with their dark blue boundaries extending approximately $0.7\lambda_0$ horizontally and $0.6T_0$ vertically. These two regions were previously identified for the weak breaker in figure 2 of Erinin *et al.* (2019). In the present paper, the rectangular regions encompassing each of the main droplet producing regions are called Region I (roughly the lower half of each plot and consisting of Regions I-A (orange background) and I-B (blue background) as previously defined in figure 3) and II (roughly the upper half of each plot, tan background). To aid in comparing the droplet production regions with the surface profile data shown in Part 1, the $N(\tilde{x}, \tilde{t})$ contour plots from figure 5 for the three waves are superposed on the contour plots of the average surface height ($\langle y - y_c^i \rangle$) in figures 6(*a*) - (*c*), the profile normal distance standard deviation ($n_{sd}$) in figures 6(*d*) - (*f*) and the profile arc length standard deviation ($s_{sd}$) in figures 6(*g*) - (*i*). The original contour plots of $\langle y - y_c^i \rangle$, $n_{sd}$ and $s_{sd}$ are given in figures 11, 12 and 13, respectively, of Part 1 and the definitions of $n_{sd}$ and $s_{sd}$ can be found in figure 8 of Part 1 and in the caption of figure 6 in the present paper. As can be seen in figures 6(*a*) - (*c*), the area of large droplet production in Region I is generally aligned with the breaking wave crest and in Region II with the following wave crest.

In Region I, the local maxima of *N* vary in spatio-temporal location and in magnitude for the three breakers. From inspection of white-light movies of the breaking events and the comparisons with the contour plots of $n_{sd}$ and $s_{sd}$, these maxima seem to be associated with different physical surface processes in the breakers. For the weak breaker, there are three prominent local maxima in Region I, as originally identified and discussed in detail in Erinin *et al.* (2019). The first local maximum (identified by the white filled black circle in figure 5 (*a*)) occurs shortly after and downstream (to the left) of jet impact (at ($\tilde{x}, \tilde{t}$)= (0, 0)). This location is near the end of the first indentation, which is marked by the rightmost magenta curve at the bottom of the plot and by the rightmost string of downward magenta triangles in figure 3(*a*). The second maximum, marked by the white-filled black triangle, is just to the right of the second indentation and occurs in a region of low surface normal and arc length standard deviations between two regions of high standard deviation which stem from the first and second splashup on the two sides of the second indent, see figures 6(*d*) and (*g*). The third local maximum (marked by the solid black square) is located close to the region of high standard deviation associated with the bursting of large air bubbles that were initially entrained under the plunging jet. This local maximum is on the back face of the wave crest, figure 6(*d*) and (*g*).

For the moderate and strong breakers, Region I contains only two prominent local maxima which are in similar spatio-temporal locations for the two waves. The first maxima (marked by the white filled black circle) is located at approximately ($\tilde{x}, \tilde{t}$) = ($0.220\lambda_0, 0.259 f_0^{-1}$) and ($0.212\lambda_0, 0.374 f_0^{-1}$) for the moderate and strong breakers, respectively. This region seems to issue from the end of the first indentation and droplets produced in this region are confined to a narrow spatial and temporal location. The number of droplets increases with breaker intensity. The second local maxima (marked by the white filled black triangle) is located on or to the right of the second indentation, at approximately ($\tilde{x}, \tilde{t}$) = ($0.381\lambda_0, 0.374 f_0^{-1}$) and ($0.415\lambda_0, 0.518 f_0^{-1}$) for the moderate and strong breakers, respectively. As in the second maximum for the weak breaker, these maxima occur in a region of low surface normal and arc length standard deviation which is enclosed by two regions of high surface normal and arc length standard deviation located directly downstream and upstream, see figures 6(*e*), (*f*), (*h*), and (*i*). From visual inspection of white-light movies, this local maximum is associated with the splash region at the leading edge of the breaking zone and the sudden eruption of large air bubbles that were entrapped under the plunging jet at impact. Thus, it appears that

14 *M. A. Erinin, C. Liu, S. D. Wang, X. Liu and J. H. Duncan*14 *M. A. Erinin, C. Liu, S. D. Wang, X. Liu and J. H. Duncan*

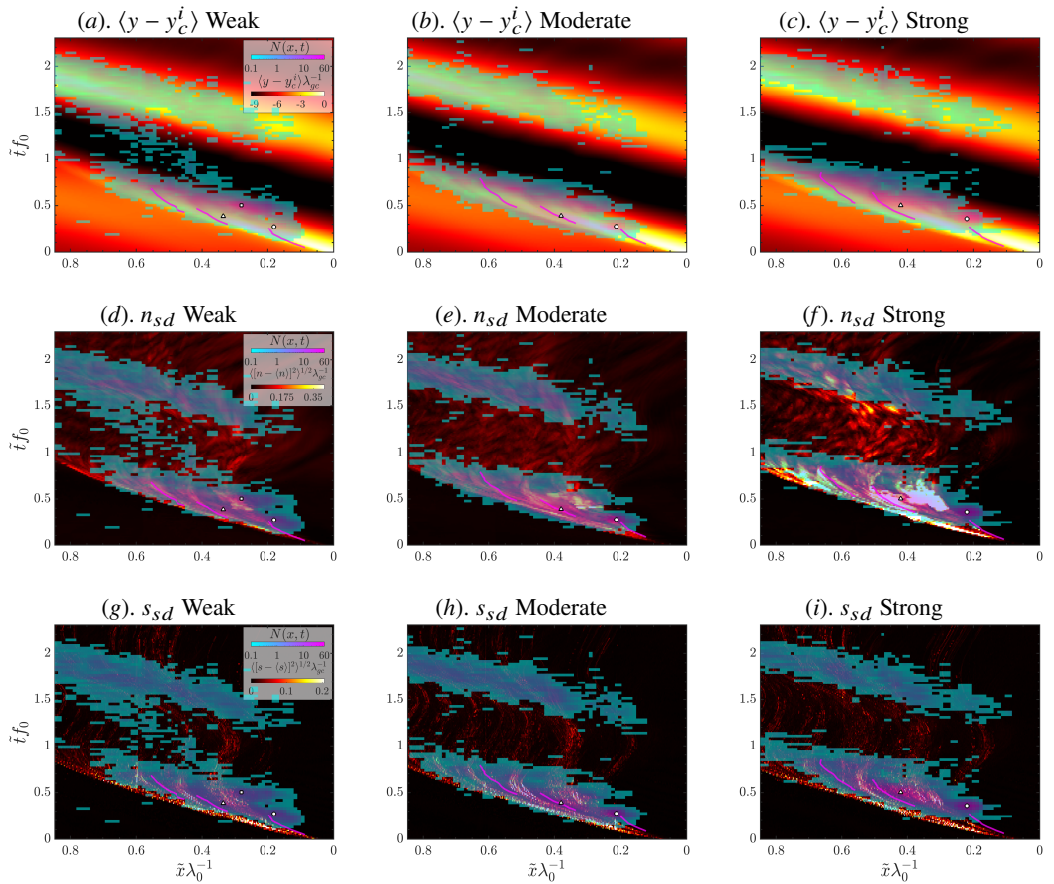

Figure 6: Contour plots of $N(\tilde{x}, \tilde{t})$ as 50%-transparent overlays on contour plots of the ensemble average surface height ($\langle y - y_c^i \rangle$, top row), the surface normal standard deviation ($n_{sd}$, middle row) and the arc length standard deviation ($s_{sd}$, bottom row) for the three breakers with plots for the weak, moderate and strong breakers in the left, middle and right columns, respectively. As defined in Part 1, $n_{sd} = \sqrt{\langle [n - \langle n \rangle]^2 \rangle}/\lambda_{gc}$, where $n$ is the distance between the average profile and an individual profile measured at each point along the average profile in the direction of its local normal and $s_{sd} = \sqrt{\langle [s/s_m - 1]^2 \rangle}$, where $s$ is the local arc length of an individual profile and $s_m$ is the corresponding local arc length of the ensemble average profile. In each plot, the $N(\tilde{x}, \tilde{t})$ overlay contours are presented in the teal-pink color range. The $\langle y - y_c^i \rangle$, $n_{sd}$ and $s_{sd}$ contours in the background are opaque and are plotted in the black - yellow color range. The three short curves in the lower portion of the plots, which are colored magenta in the underlying plots and appear red in the top row of plots, are the three indentations as described in Part 1, figures 9 and 11 to 13. The circles, triangles and squares with black outlines and white fill are the same points as are in the corresponding $N(\tilde{x}, \tilde{t})$ plots in figure 5.

the second maxima in the moderate and strong breakers is composed of droplets from the leading edge splashing and large bubble bursting that create the second and third maxima, respectively, in the weak breaker.

In Region II, there are no pronounced local maxima and the magnitudes of $N(\tilde{x}, \tilde{t})$ are similar for the three waves. Observations from the white-light movies indicate that the droplets measured in Region II are primarily the result of small bubbles that burst when reaching the free surface, after the main sources of droplets on the breaking crest have ceased production. It is shown in § 3.4 that these small bursting bubbles are generated with comparatively low vertical velocities. Thus, it is theorized that the reason that the droplets in Region II are measured only over the following wave crest, is that the droplets do not travel vertically more than a few centimeters and only the crest of the following wave is within this distance from the measurement plane.

| Breaker Type | Weak | Moderate | Strong |
|---|---|---|---|
| Region I-A | | | |
| $x$-range/$\lambda_0$ | 0 - 0.167 | 0 - 0.260 | 0 - 0.260 |
| $t$-range/$T_0$ | 0 - 0.288 | 0 - 0.320 | 0 - 0.403 |
| $N_{I-A}$ | 228 (34.7) | 235 (28.0) | 388 (34.6) |
| Region I-B | | | |
| $x$-range/$\lambda_0$ | 0.209 - 0.922 | 0.260 - 0.922 | 0.260 - 0.922 |
| $t$-range/$T_0$ | 0.316 - 1.150 | 0.320 - 1.150 | 0.403 - 1.150 |
| $N_{I-B}$ | 283 (43.1) | 369 (44.0) | 565 (50.4) |
| Region II | | | |
| $x$-range/$\lambda_0$ | 0 - 0.923 | 0 - 0.923 | 0 - 0.923 |
| $t$-range/$T_0$ | 1.150 - 2.300 | 1.150 - 2.300 | 1.150 - 2.300 |
| $N_{II}$ | 146 (22.2) | 235 (28.0) | 169 (15.0) |
| $N$ | 657 | 839 | 1122 |

Table 1: The number of droplets generated per breaking event per meter of crest length in Regions I-A, I-B and II as well as the total for each of the three breakers. The spatial and temporal limits of the three droplet producing regions are also given. The percent contributions of each region relative to the total number of droplets is given in parenthesis.

In order to make better quantitative comparisons of $N(\tilde{x}, \tilde{t})$ from one breaker to another, the distribution is integrated in $\tilde{x}$ to obtain $N(\tilde{t})$ and in $\tilde{t}$ to obtain $N(\tilde{x})$. In addition, the integration

$$N'(\tilde{x}') = \int_0^{2000 \text{ ms}} N(\tilde{x}', \tilde{t}) \, d\tilde{t} \qquad (3.1)$$

is performed where $\tilde{x}' = \tilde{x} + \langle u_c^i \rangle \tilde{t}$ is the streamwise coordinate of a reference frame moving with speed $\langle u_c^i \rangle$, the speed of the crest point at the moment of jet impact, see Part 1 for details. The results are shown in figure 7 where $N(\tilde{t})$, $N(\tilde{x})$ and $N'(\tilde{x}')$ are plotted in subplots (*a*), (*b*) and (*c*), respectively. Each subplot contains three curves, one for each breaker. See the figure caption for details.

The three curves of $N(\tilde{t})$ and $N(\tilde{x})$, in subplots 7(*a*) and 7(*b*), respectively, where jet impact occurs at the left and right ends of the horizontal axes, respectively, contain a number of local maxima. The locations of the first of these maxima (moving to the right in (*a*) and to the left in (*b*)) are marked by solid red vertical lines and also by vertical and horizontal lines on the edges of the the contour plots of $N(\tilde{x}, \tilde{t})$ in figure 5. In the curve for the weak breaker in subplot 7(*b*), the overall maximum of the curve is marked as the first local maximum since the other local maxima are not significant. As can be seen in the contour plots of figure 5, the two solid red line segments would cross near the white filled black circles, indicating that these $N(\tilde{t})$ and $N(\tilde{x})$ local maxima are due to droplets generated during the collapse of the crater at the bottom of the first indentation. From subplots (*a*) and (*b*), it can be seen that the peak values of $N(\tilde{t})$ and $N(\tilde{x})$, increase by factors of 3.8 and 3.9, respectively, from the weak to the strong breaker. The second local maxima are marked by vertical red dashed lines and these maxima point to the white filled black triangle in figure 5, indicating that these local maxima are due to a combination of droplets from the splash region and the bursting of large bubbles entrained under the plunging jet. The peak values of $N(\tilde{t})$ and $N(\tilde{x})$ at these locations increase by factors of 2.5 and 3.5, respectively, from the weak to the strong breaker. In the plot of $N(\tilde{t})$, there is a small local peak at about $\tilde{t} = 0.58 f^{-1}$ in all the curves and finally a consistent small peak in all curves at $\tilde{t} = 1.5 f^{-1}$, which is the midpoint of the passage of the following wave crest. This latter peak does not appear in the plots of $N(\tilde{x})$ because the integrations in time smear the effects of the breaking and following crests.

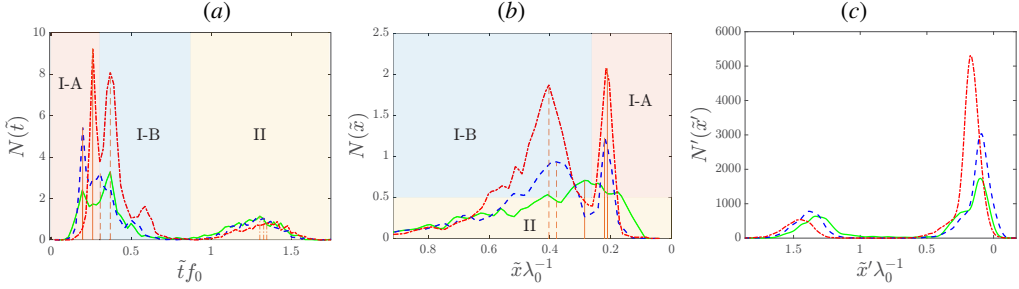

Figure 7: Droplet number distributions $N(\tilde{t})$, $N(\tilde{x})$ and $N'(\tilde{x}')$ are given in subplots (a), (b) and (c), respectively. The spatial and temporal resolution of the subplots are $\Delta x = 21.6$ mm and $\Delta t = 25.0$ ms respectively. Data for the weak, moderate and strong breakers are given as the green solid, blue dashed and red dotted lines, respectively. The colored backgrounds in subplots (a) and (b) indicate the approximate temporal and spatial limits, respectively, of the three droplet producing regions defined in Section 3.2. The function $N'(\tilde{x}')$ presented in subplot (c) is obtained by integrating $N(\tilde{x}', \tilde{t})$ over all time, where $\tilde{x}' = (\tilde{x} + \langle u_c^i \rangle \tilde{t})$, i.e. the streamwise position in a coordinate system moving downstream with speed $\langle u_c^i \rangle$, the speed of the crest point at the time of jet impact.

In the crest-fixed coordinates of subplot 7(c), the computation of $N'(\tilde{x}')$ results in a plot consisting of only two very clearly defined maxima for each breaker. This is because in this moving coordinate system, the integrations over the breaking and following crests are entirely separated and within each crest the various peaks overlap when integrating in $\tilde{t}$. The areas under the first and second peaks are the total number of drops measured in regions I and II. The magnitude of the first peak increases monotonically by a factor of approximately 3.0 from the weak to the strong breaker while the magnitude of the second peak increases from the weak to the moderate breaker and then decreases for the strong breaker to a value less than that for the weak breaker. The origin of this non-monotonic behavior is under investigation.

The number of droplets produced in each droplet producing region (I-A, I-B, and II) and the total number of droplets measured for each breaker are given in table 1 along with the precise definitions of each region. These region definitions are set according to the local minima in the plots of $N(\tilde{x})$ and $N(\tilde{t})$ in figures 7(a) and (b). As indicated in the table, the droplets generated from the closing of the first indentation crater (Region I-A) comprise on average 32% of the total produced by each breaker. Both the number of droplets in II-B and the number in II-B as a percent of the total number of droplets increase with breaker strength. The total number of droplets measured over the breaking crest, i.e., those in Regions I-A and I-B combined, comprise approximately 80% of the totals. The number of droplets produced by Regions I-A, I-B and II, and the total are plotted versus the area under the plunging jet's upper surface at the moment of jet impact, $Q^i$, and the vertical component of the crest point velocity at moment of jet formation, $\langle v_c^f \rangle$, (see Part 1 for detailed definitions) in figure 8, subplots (a) and (b), respectively. The fact that the two plots look similar is a result of the nearly linear relationship between $Q$ and $v_c^f$ as shown in Part 1.

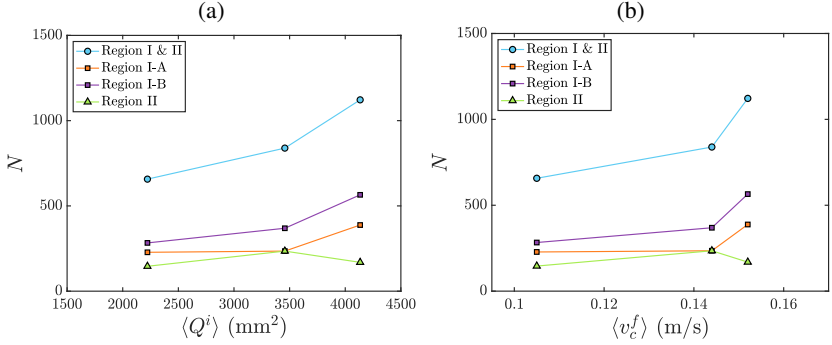

Figure 8: In subplots (*a*) and (*b*), the total number of droplets and the number of droplets measured in each subregion are plotted versus versus $Q^i$, the area under the plunging jet at $t = \langle t^i \rangle$, and $v_c^f$, the vertical velocity of the crest point at $t = \langle t^f \rangle$, respectively. As defined in Part 1, quantities with superscript $f$ were measured at the moment of jet formation and those with superscript $t$ were measured at the moment of jet impact.

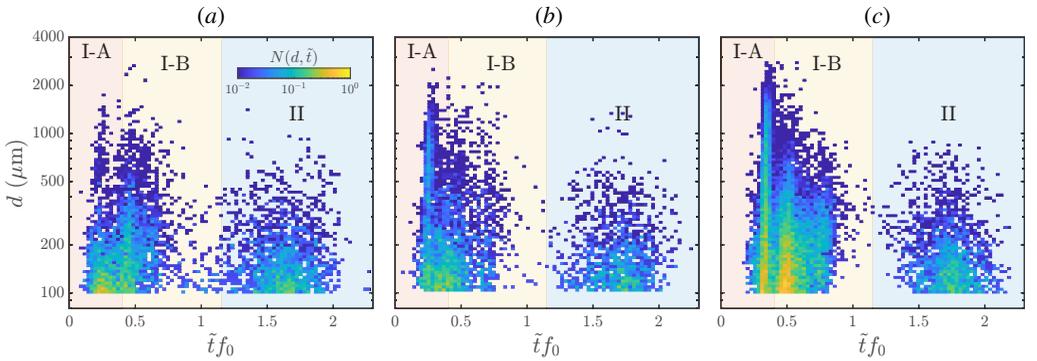

Figure 9: Contour maps of $N(d, \tilde{t})$, the number of droplets per breaking event per meter of crest length per logarithmic diameter bin, for the weak (subplot *a*), moderate (subplot *b*) and strong (subplot *c*) plunging breakers. The contour maps begin at the time of jet impact, $\tilde{t} = 0$, and cover approximately two wave periods ($\tilde{t} f_0 \approx 2.2$), with a temporal resolution of $\Delta t = 28.8$ ms. The scale of the vertical axis is logarithmic. The colored red/yellow/blue backgrounds in each subplot show the temporal limits of the three droplet producing regions as identified in § 3.2.

### 3.3. *Droplet Diameter Distributions*

The droplet diameter distributions $N(\tilde{t}, d)$ and $N(d)$ are discussed in this subsection. Contour plots of $N(\tilde{t}, d)$, for the weak, moderate and strong breakers are shown in figure 9(*a*), (*b*) and (*c*), respectively. The three contour maps look qualitatively similar with two main droplet producing regions, I and II, and separate peaks in $N$ in Regions I-A and I-B at small $d$ (as seen most clearly in subplot (*c*)). There are also a number of large scale trends with varying breaker intensity. First, the number of droplets generally increases with breaker intensity at all diameters. Second, the diameters of the largest droplets ($d = d_{max}$), which are produced early in the breaking events as the crater at the bottom of the first indentation closes at $\tilde{t} \approx 0.41 f_0^{-1}$, increase from $d_{max} \approx 1200$ µm for the weak breaker to $d_{max} \approx 3000$ µm for the strong breaker (note the logarithmic scale of the $d$ axis in the plots). Finally, the ratio of the maximum diameters in Region I-A to those in I-B increases with increasing breaker intensity.

The droplet diameter distributions, $N(d)$, are presented in the five log-log subplots in diameter. This is a classic problem in bubble and droplet measurements and to emphasize this inadequacy the power laws fitted to the large diameter droplet data are drawn as dashed lines. The values of $\alpha$, $\beta$ and $d_i$ for the three breakers studied herein are given in table 2. The slope for the small diameters, $\alpha$, increases monotonically with increasing breaker strength while the slope for the large droplets first decreases and then increases. The intersection diameter increases monotonically with values ranging from 820 to 1480 µm. If the data from subplot (*a*) is plotted with the $d/d_i$ as the independent variable, this normalization nearly collapses the data to a single two-slope curve.

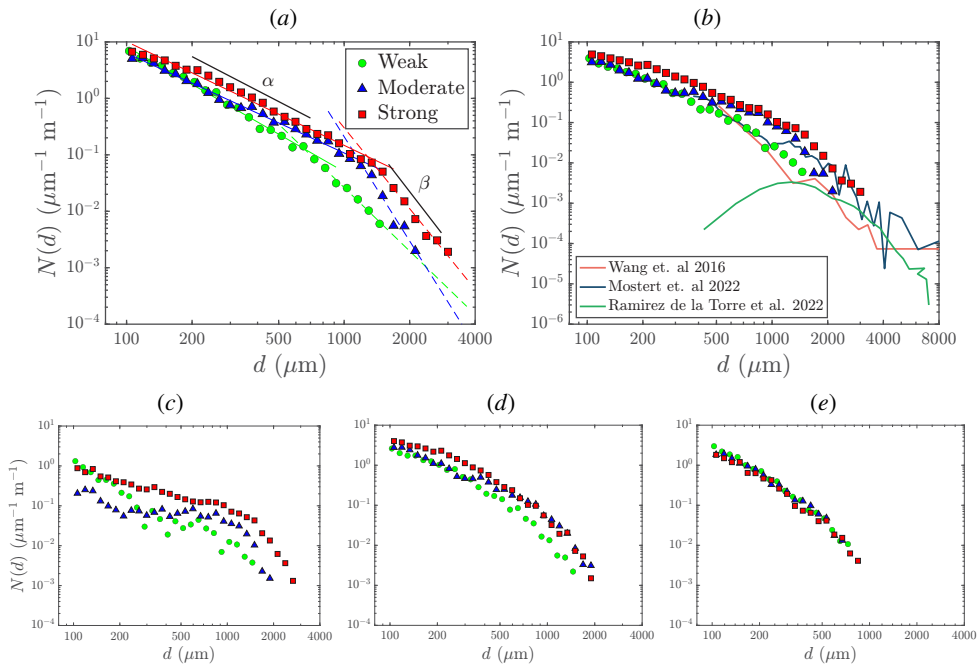

Figure 10: Plots of the droplet diameter distributions, $N(d)$, for various regions of the measurement plane. In each plot, the green circles, blue triangles, and red squares are the data from the weak, moderate, and strong breakers, respectively. Along the horizontal axis, the diameter bins are logarithmically spaced from $d = 100$ to $4000$ μm with 32 bins. In order to decrease the noise in these plots, all bins containing less than five measured droplets (all found at large $d$) are removed. The data in subplot (*a*) are from the droplets measured over the full range of time ($0 < t < 2200$ ms) and streamwise position ($0 < x < 1050$ mm). Subplot (*a*) shows $N(d)$ and the dashed lines are fits of straight lines to the values from the large and small diameter ranges as discussed in the text. The solid lines are the results from numerical simulations and experiments, see plot key. Subplot (*b*) shows $N(d)$ for Region I along with $N(d)$ from three other studies. See the text for details. In subplots (*c*), (*d*) and (*e*), $N(d)$ for Regions I-A, I-B, and II, respectively, are shown. The spatio-temporal limits of these regions are defined in table 1.

figure 10. Each subplot contains three data sets from the present experiment, one for each breaker. See the figure caption for additional details. Subplot 10 (*a*) contains the data for the entire data set for each of the three breakers. The curves are a bit noisy but generally indicate an increasing droplet number with increasing breaker strength at diameters above approximately 400 μm, the curves converge at smaller diameters. A number of authors have measured droplet diameter distributions in breaking wave and wind wave systems in the laboratory (see for example Ramirez de la Torre *et al.* (2022); Erinin *et al.* (2022); Veron *et al.* (2012); Ortiz-Suslow *et al.* (2016)), the field ((Wu *et al.* 1984; Monahan 1968)), and in direct numerical simulations with and without wind ((Wang *et al.* 2016; Tang *et al.* 2017*b*; Mostert *et al.* 2022)). Many of the distributions in these studies are either fitted or compared to separate power law functions at small and large droplet diameter ranges. In view of these earlier studies, straight lines were fitted by least square error minimization separately to the smaller diameter and larger diameter droplet data in subplot (*a*) for each of the three waves. The droplet diameter ($d_i$) at the boundary between the larger and smaller diameter groups was determined by an iterative bisection-like routine outlined in Erinin (2020) and the supplemental material in Erinin *et al.* (2019). It should be noted that this determination of $d_i$ is inherently inaccurate because of the small number of droplets in each bin at the larger

Figure 10(*b*) contains the present data from Region I and the droplet diameter distributions from the DNS results of Wang *et al.* (2016) and Mostert *et al.* (2022) as well as the no-wind case of breaking as a focused wave packet approaches a shoal in the laboratory experiments of Ramirez de la Torre *et al.* (2022). Information about the numerical studies is given at the end of § 1 in the present paper, as well as in the original references. The data from Mostert *et al.* (2022) is obtained from the $Bo = 1,000$ case in their figure 19 and the capillary length scale $\ell_c$ is taken as 2.73 mm. The data from the no-wind $ak = 0.66$ case in figure 7 of Ramirez de la Torre *et al.* (2022) is reproduced by taking the value of the average droplet diameter as $\overline{D_e} = 1.75$ mm. The data from the high resolution case in figure 13 of Wang *et al.* (2016) is reproduced directly. In the cases of Wang *et al.* (2016) and Ramirez de la Torre *et al.* (2022), the curves are adjusted vertically so that they approximately match the present data. The methods used to count droplets in each paper are different from those used in the present study and this adjustment is necessary in some cases. As can be seen from the subplot, the shapes of the two CFD-based probability distributions are fairly similar to the present results, in spite of the differences in the wave motion, path to breaking, and wavelength between the DNS and the present study. Distributions from the experiments of Ramirez de la Torre *et al.* (2022) fall off the present distributions at small diameter.

The droplet diameter distributions ($N(d)$) for droplet producing regions I-A, I-B and II in the present experiments are given in subplots (*c*), (*d*) and (*e*) of figure 10, respectively. Each plot is created from many fewer droplet measurements than in subplots (*a*) and (*b*) but still displays interesting features. In Region I-A, the two linear regions of the curves for the moderate and strong breakers are still evident; however, in view of the noisy data, the power-law fitting was not performed. The vertical separation between the curves is larger than in the distribution for the full data set, but the trend with breaker strength is only monotonic for $d \gtrsim 300$µm. For the data in Region I-B, the curves are less noisy and closer together. The two-slope nature of the previously discussed distributions has be replaced by smooth arcs in which the local slope magnitude increases with increasing $d$. This trend continues in Region II where the data forms smooth curves that are nearly on top of one another. The structure of these data sets is thought to be a result of the different droplet generation mechanisms in the three regions. The very different curves in Region I-A are thought to be a result of variations in droplet production during the crater collapse at the bottom of the first indent as was discussed in § 3.2 and as seen in Movie 1 given as supplementary material. This hypothesis is also supported by the droplet velocity data reported in the following subsection. In Region II, droplets are generated as small bubbles come to the surface and burst. This leads one to hypothesize that the main difference between the three distributions in figure 10(*e*) is the number of droplets, $N_{II}$, i.e., the area under the three curves. Though the three data sets look similar in the log-log plot, $N_{II}$ varies between the data sets, see table 1. As a test, the probability distribution, $N(d)/N_{II}$, was examined (not presented herein). In this PDF plot, the three data sets in are even closer together than in the $N(d)$ plot, thus supporting this small-bubble-bursting hypothesis. Finally, comparing the $N(d)$ plots in the three sub regions to that in the entire data sets in subplot (*a*) indicates that Region I-A is the only region in which the data sets show the two-slope structure found in the full data sets. This supports the idea that this two slope structure is create by the droplets generated by the closure of the first indentation.

| Breaker Type | Weak | Moderate | Strong |
|---|---|---|---|
| All Regions | | | |
| $\alpha$ | -2.4 | -2.1 | -1.9 |
| $\beta$ | -5.4 | -6.1 | -5.8 |
| $d_i$ (μm) | 820 | 1140 | 1480 |

Table 2: Parameters for the straight lines fitted by least squares error minimization to the distributions of droplet diameter in figure 10(*a*). The variables $\alpha$ and $\beta$ are the slopes of the straight lines for the small and large diameter droplet data, respectively, while $d_i$ is the diameter at which the two lines intersect.

### 3.4. *Droplet Velocity Distributions*

The distributions of the streamwise and vertical droplet velocity components, $u$ and $v$, respectively, and the 2D speed, $V = \sqrt{u^2 + v^2}$, are presented and discussed in this subsection. In all cases, $u$, $v$, and $V$ are reported relative to the laboratory reference frame. Contour plots of the droplet number distributions $N(u, d)$, $N(v, d)$ and $N(V, d)$ are given for all the droplets for each of the three breakers in the nine subplots of figure 11, see caption for details. From these contour plots, one can see that as the breaking intensity increases (moving down from plot to plot in each column), the number of droplets with large diameter and the range of $u$, $v$ and $V$ also increase. The peaks of the distributions are located at small diameters and at small values of $u$, $v$ and $V$. The distributions $N(u, d)$ (left column) indicate that there are droplets with positive (in the direction of wave travel) and negative horizontal velocities at all diameters and that the distributions are centered vertically at slightly positive values of $u$. The distributions $N(v, d)$ (center column) are located, of course, entirely in the region of positive $v$ with a few droplets having speeds of nearly 4 m/s and the number of these fast moving droplets increasing with breaker intensity. In considering these droplet speeds, it should be kept in mind that for these breakers the plunging jet speed ($\langle V_j^i \rangle$) and crest speed ($\langle u_c^i \rangle$) at jet impact are approximately, 2.0 m/s and 1.3 m/s, respectively, (see table 3 in Part 1 for details). The droplet velocity estimates divided by the crest speeds from the numerical calculations reported in Mostert *et al.* (2022), where the breaker wavelengths are approximately 30 and 50 cm, are similar in range to the present results where the nominal wavelength is 118 cm.

Probability distributions of $u$, $v$ and $V$ are presented for all of the droplets and for the droplets in Regions I-A, I-B and II in figure 12, subplots (*a*) through (*l*). See the figure caption for additional details. Average values of $u$, $v$ and $V$ corresponding to these plots are given in table 3. The PDFs of $u$ (the four subplots in the left column) are approximately symmetric with peaks located between approximately 0.1 and 0.3 m/s and average values between 0.05 and 0.75 m/s. Among the three sub-regions, the average values of $u$ in Region I-B (0.58 m/s averaged over the three breakers) are the highest while the average $u$ in Regions I-A and II are only 0.18 and 0.21 m/s, respectively. As mentioned above, droplets in Region II are generated as small bubbles reach the surface and burst. Small bubbles bursting on a water surface with zero mean flow are expected to produce symmetric $u$ (and $w$ if measured) PDFs with zero mean. Thus, the small positive mean and most probable $u$ values of approximately $u = 0.14 \langle u_c^i \rangle$ in Region II are probably an indication of horizontal fluid motion due the decaying surface drift current from breaking and the particle velocity in the crest of the nonbreaking following wave. In the PDFs of $v$ (the four subplots in the middle column), the average value of $v$ and the range of $v$ decrease monotonically in going down in the column of plots from all droplets to Regions I-A to I-B to II. These differences and trends in the $u$ and $v$ PDFs between the three droplet producing regions and the three breakers are consistent with the picture of the droplet producing mechanisms in the three regions: crater collapse in the first indentation in Region I-A, splashing and large and small bubble bursting in Region I-B and small bubble bursting in Region II. The corresponding pdfs for 2D droplet speed, $V$, are given for the interested reader.

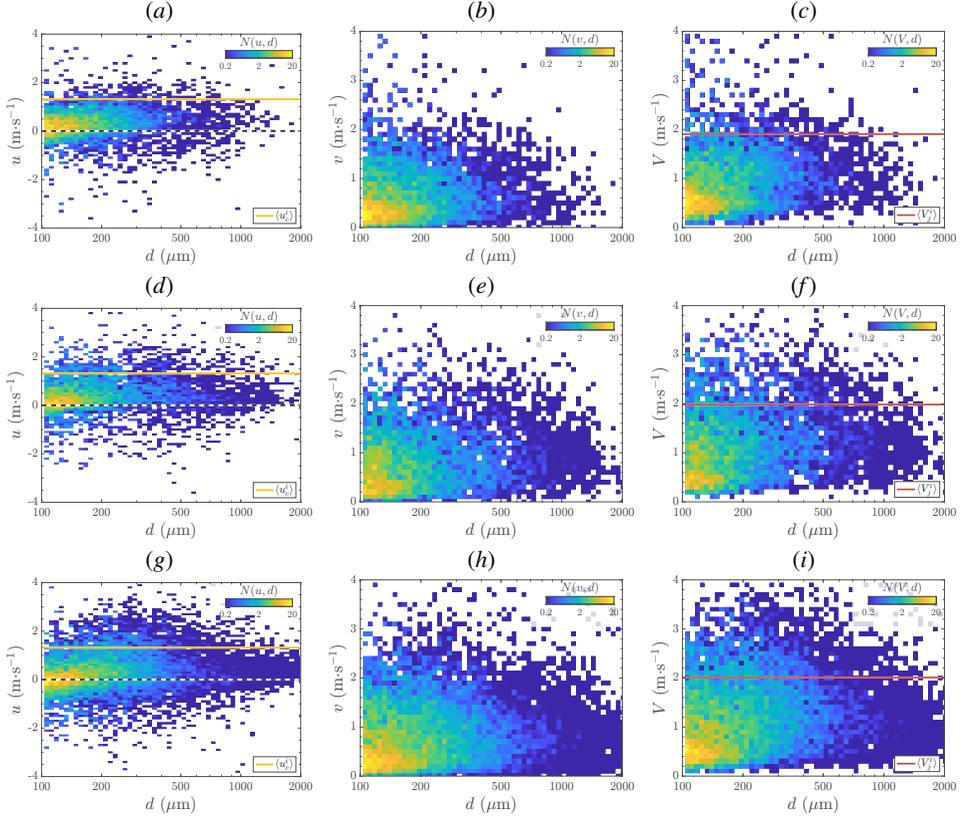

Figure 11: Contour maps of the number of droplets per breaking event per meter of crest width per bin widths as a function of droplet velocity component $u$ and $d$, $N(u,d)$ (left column), $N(v,d)$ (middle column), and $N(V,d)$ (right column), for the weak (top row), moderate (middle row), and strong (bottom row) plunging breakers, where $V = (u^2 + v^2)^{0.5}$ is the 2D speed. In all subplots, the vertical axis has equally spaced bins of height 0.1 m/s and the horizontal axis has 80 logarithmically spaced bins with bin edges from $d = 100$ on the left to 2844 µm on the right. The horizontal solid red and yellow lines in the $N(V,d)$ and $N(u,d)$ subplots, respectively, are located at the values of $\langle V_j^i \rangle$ (the plunging jet speed at impact) and $\langle u_c^i \rangle$ (the crest point speed at jet impact), respectively, from table 2 of Part 1.

The differences in the droplet velocities and their relationship to the droplet production mechanisms is further demonstrated by the examination of the droplet velocity direction, as measured by the angle ($\theta$) between the 2D velocity vector and horizontal direction of wave motion. The distributions of $N(d, \theta)$ and $N(\theta)$ are presented in seven polar subplots in figure 13. See the figure caption for details. Subplots (*a*), (*b*) and (*c*) are contour plots of $N(d, \theta)$ for all droplets in the weak, moderate and strong breakers, respectively. The radial and azimuthal coordinates are $d$ and $\theta$, respectively. All three distributions have an isolated region of high $N(d,\theta)$ at small $100 < d \leqslant 300$ µm and $\theta \approx 60°$. To better compare the three waves, the distribution $N(\theta)$ for all the droplets in each wave is plotted in polar coordinates with $N(\theta)$ and $\theta$ as the radial and azimuthal coordinates, respectively, in subplot (*d*). There is one curve for each breaker intensity. The curves are somewhat irregular but one can see that the value of $\theta$ at the highest values of $N(\theta)$ increase with increasing breaker intensity. This increase in $\theta$ with breaker intensity is reflected in the average values of $\theta$ for the all droplets as given in the next to last row of table 3. Subplots (*e*), (*f*) and (*g*) contain polar plots of $N(\theta)$ for Regions I-A, I-B and II, respectively. These plots and the $\bar{\theta}_m$ data in table 3 demonstrate clear differences in droplet production between the three regions. In the three subplots, one can see that $\theta_m$ is only about 20° downstream from vertical, i.e., in Regions I-A and II, while it is on the order for 40° downstream from vertical in Region I-B. This trend is reflected in the average values of $\theta$ in the three regions: 81.1°, 56.2° and 74.1° in Regions I-A, I-B and II, respectively, where each value is the average over the three waves. The narrow distribution of nearly vertical motion of the droplets in Region I-A is clearly observed in Movie 1 given as supplementary material and is thought to be related to the generation during the closure of the crater at the bottom of the first indentation. The narrow nearly vertical distribution in Region II is probably due to a combination of the distribution of droplet velocity in an individual small bubble bursting event and the fact that measuring the droplets as they pass through the measurement plane above the generation point favors droplets generated with velocity vectors pointing up.

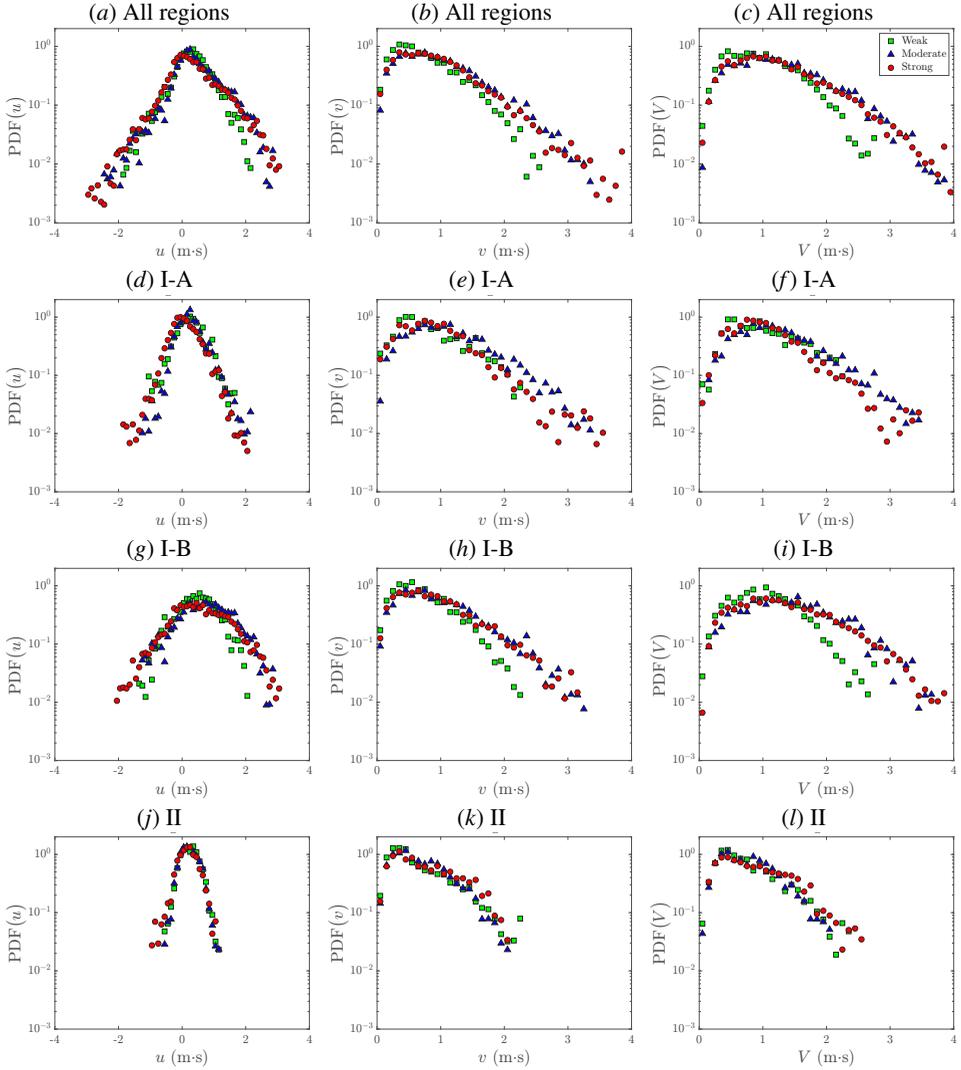

Figure 12: The probability density functions (PDFs) of the horizontal velocity component (*u*, left column), and vertical velocity component (*v*, middle column) and the 2D speed ($V = \sqrt{u^2 + v^2}$, right column) of droplets as they pass through the measurement plane. The data covering all regions and Regions I-A, I-B and II are given in the top through bottom rows, respectively. The width of all bins on the horizontal axis is 0.1 m/s in all subplots and the PDF calculations are terminated after a bin contains less than 3 droplets. In each plot, data is given for the three breakers and the plotting symbol definitions are given in the key in plot (*c*).

| Run | All regions | | | Region I-A | | | Region I-B | | | Region II | | |
|---|---|---|---|---|---|---|---|---|---|---|---|---|
| | W | M | S | W | M | S | W | M | S | W | M | S |
| $\bar{u}$ (m/s) | 0.31 | 0.34 | 0.22 | 0.25 | 0.23 | 0.05 | 0.48 | 0.75 | 0.50 | 0.24 | 0.21 | 0.18 |
| $\bar{v}$ (m/s) | 0.63 | 0.87 | 0.82 | 0.74 | 1.05 | 0.85 | 0.63 | 0.84 | 0.81 | 0.53 | 0.62 | 0.67 |
| $\bar{V}$ (m/s) | 0.91 | 1.17 | 1.09 | 0.94 | 1.16 | 0.96 | 1.00 | 1.38 | 1.25 | 0.67 | 0.74 | 0.81 |
| $\bar{\theta}$ (deg.) | 65.6 | 66.5 | 75.0 | 78.4 | 75.4 | 89.5 | 54.3 | 51.0 | 63.3 | 69.0 | 73.2 | 80.0 |
| $N$ | 657 | 839 | 1122 | 228 | 235 | 388 | 283 | 369 | 565 | 146 | 235 | 169 |

Table 3: The average horizontal velocity component ($\bar{u}$), vertical velocity component ($\bar{v}$), 2D speed ($\bar{V} = \sqrt{\bar{u}^2 + \bar{v}^2}$), 2D average velocity angle ($\bar{\theta}$) relative to horizontal ($\bar{\theta} = 0$ defined as downstream) and number of droplets ($N$) are given for each wave and separately for Regions I-A, I-B, and II. The values $N$ are the number of droplets per breaking event per meter of crest width in each region, the same numbers are reported in table 1.

### 3.5. *Low-order Predictions of Droplet Generation Sites and Time of Flight*

As discussed above, the droplet data presented herein is measured as the droplets move upward through the measurement plane just above the wave crest. In an effort to determine an approximate time and position of the generation of each droplet at the water surface, the measured droplet data is used with a model of the droplet motion to simulate the trajectories of the droplets backward in time until they intersect with the time evolving ensemble average profile history. This model also yields an estimate of the droplet time of flight ($\Delta t_f$) as it travels from the water surface to the measurement plane. Estimates of $\Delta t_f$ are essential for caculations of the reduction in droplet diameter due to evaporation before each droplet is measured. In this model, the forces on the particles are assumed to be those due to gravity and drag relative to still air, with the drag coefficient taken from the correlation for a spherical particle given by Cheng (2009). Since, the air motion is very likely an important factor in determining the droplet motion, particularly for the smaller droplets, the degree to which the computed generation sites seem plausible is used to indicate the importance of the air velocity in the calculation and thus motivate additional research.

The results of these droplet trajectory calculations are shown in figure 14 where two plots of the ensemble averaged profiles from the strong breaker are shown along with the same set of droplets measured from three realizations of the strong breaker. In subplot (*a*), which is a copy of the plot in figure 3(*c*), the droplets are shown on the profiles measured at the time the droplets crossed the measurement plane and at the corresponding streamwise positions. In subplot figure 14(*b*), the water surface profiles in (*a*) are repeated but the droplets are plotted on the profiles and at streamwise positions computed by the backward tracking model. As can be seen by comparing the two plots, many of the droplets that were measured over the breaking crest in Region I-B are predicted to have been generated near where they were measured, i.e., on the turbulent breaking region. Also, the area where many droplets were measured in Region I-A in a sharply defined region filling the region between profiles (*ii*) and (*iii*) in plot (*a*) have tightened to a smaller range of time just after profile (*ii*). The number of droplets in this region of (*b*) is only a little less than in the corresponding region of (*a*); this result is not visible in the plots because many of the droplets are plotted on top of one another. An additional important feature of the plots is that many of the smaller droplets and some larger ones, 40 percent of the total of 2,158 droplets in the figure, are predicted to have been generated on the smooth water surface downstream (to the left) of the leading edge of the breaking zone. Though examination of the white-light movies does indicate that a few secondary droplets are generated in this downstream region due to primary droplet impacts, the large number of droplets predicted to be generated is thought to be nonphysical and an indication of the need for including the temporally evolving breaker-generated air flow field in the calculation of the droplet motion.

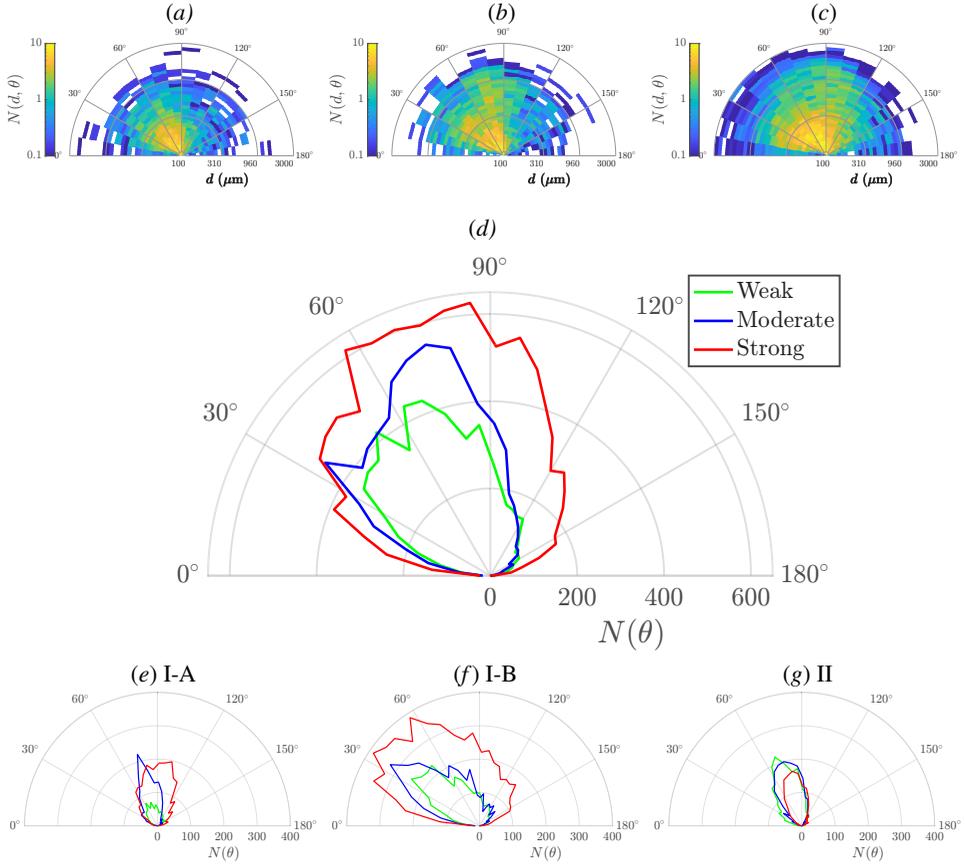

Figure 13: The subplots (*a-c*) are polar contour maps of $N(d, \theta)$, the number of droplets per breaking event per meter of crest length per diameter bin per $\theta$ bin, where $\theta$ is the angle of the mean 2D velocity vector ($\vec{v} = u\hat{i} + v\hat{j}$) relative to horizontal in each diameter bin for the weak, moderate and strong breakers, respectively. The radial and azimuthal coordinates are $d$ and $\theta$, respectively. The motion of the wave is in the direction of $\theta = 0°$ and $\theta = 90°$ is vertically up. Subplots (*d-g*) are polar line plots of $N(\theta)$, the number of droplets per breaking event per meter of crest length per $\theta$ bin for the three breakers for all regions in (*d*) and Regions I-A, I-B, and II in (*e-g*), respectively. The radial and azimuthal coordinates are $N(\theta)$ and $\theta$, respectively. The $\theta$ bin width = $5.625°$ in all plots and the $d$ bins in subplots (*a-c*) are logarithmically spaced from $d = 100$ to $3000$ µm with 63 bins.

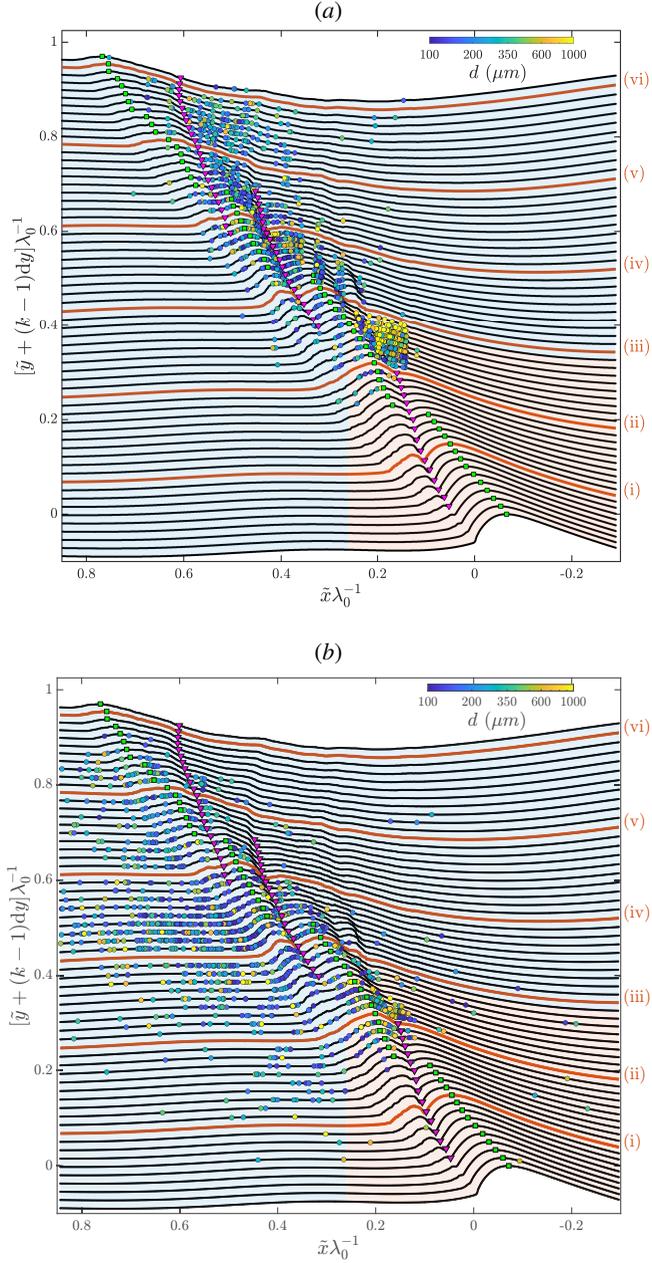

Figure 14: (*a*.) The spatio-temporal distribution of droplets from three realizations of the strong breaker plotted on the corresponding ensemble averaged wave profiles (from 10 realizations). Each droplet is plotted on the profile when and the streamwise position where it crossed the measurement plane. This is the same plot as given in figure 3(*c*)) (see that figure for more plotting details). (*b*.) The spatio-temporal positions of the droplets (*a*) projected back to their generation sites as determined by the simplified droplet motion model described in the text. Figure 3(*c*) is shown here to facilitate the comparison between the measured droplet locations in (*a*) and the estimates of the locations where they were generated in (*b*).

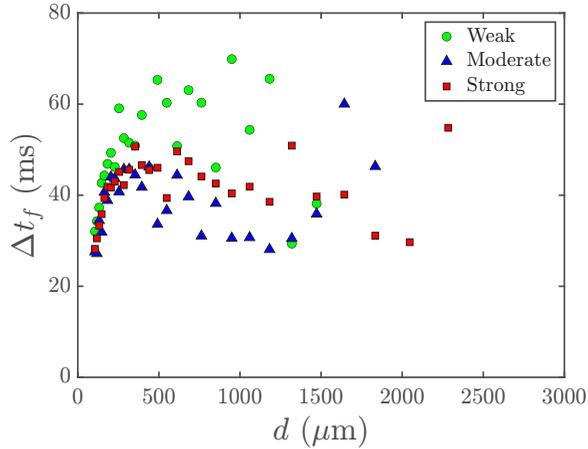

Figure 15: Estimates of the time of flight of droplets as a function of droplet diameter for the weak, moderate, and strong breakers. These estimates are based on the time, position and velocity of the droplets measured as they move upward through the measurement plane, the ensemble average profile sequence and the simplified droplet motion model discussed in § 3.5. The standard deviation of $\Delta t_f \approx 30$ ms for $d \approx 100$ µm and $\approx 50$ ms for $d \approx 500$ µm.

The calculated values of $\Delta t_f$ for the droplets considered in figure 14 are presented as a function of $d$ in figure 15. From the plot, it can be seen that the data for the droplets with the smallest diameters fall on a single line with positive slope and values ranging from $\Delta t_f \approx 25$ ms at $d = 100$ µm to $\Delta t_f \approx 40$ ms at $d = 200$ µm. For larger diameters, the data at each $d$ spreads out vertically with the smallest to largest $\Delta t_f$ values ranging from approximately 30 ms to 70 ms and generally increasing monotonically with breaker strength. These times of flight were used in § 2.3 to estimate the effect of evaporation on the measured diameters of the droplets in this study.

## 4. Conclusions

Measurements of the droplets produced by three plunging breaking waves are presented and discussed. The breakers are created from mechanically generated dispersively focused wave packets that differ primarily by only modest changes in overall amplitude. The average frequency of the wave packet is $f_0 = 1.15$ Hz (wave period $T_0 = 1/f_0 = 0.870$ s) for all breakers and this corresponds to a wavelength $\lambda_0 = 1.18$ m by linear wave theory. The breakers, which are described in detail in Part 1 of this two-part paper, are designated as weak, moderate and strong. By using two cinematic in-line holography systems (operating at 650 fps), the droplet diameters ($d \geqslant 100$ µm), their positions, and the streamwise and vertical components of their velocity are measured as the upward moving droplets cross a measurement plane located 1.2 cm above the highest point reached by the wave crest during the breaking process. The droplet measurements were recorded for a time of $2.270T_0$ and over a streamwise distance of $0.9\lambda_0$ after the time and location of plunging jet impact, respectively. The data set for each wave was obtained from approximately 140 breaking events.

It is found that for all three waves the droplets cross the measurement plane in two spatio-temporal regions, the first over the breaking crest (Region I) and the second over the following wave crest (Region II). (This result was first reported for the weak breaker in Erinin *et al.* (2019).) Approximately, 77.8, 72.0 and 84.0 percent of the droplets are measured over the breaking crest in the weak, moderate and strong breakers, respectively, while the total number of droplets measured per breaking event is $N = 657$, 839 and 1122, respectively. It is found that $N$ increases nearly linearly with $Q_i$, the area under the upper surface of the plunging jet as defined in Part 1.

In Region I, three primary droplet generation mechanisms are identified. The first mechanism is the closure of the indentation that forms just upstream of the upper surface of the plunging jet (called Region I-A herein); this mechanism contributes approximately 32% of the total number of droplets per breaking event. The second and third mechanisms are the bursting of large bubbles entrained at the moment of jet impact as they come to the surface on the back face of the wave and splashing and small bubble bursting in the turbulent zone created by the jet impact on the front face of the wave. Both of these processes occur in a Region labelled I-B and together contribute approximately 46% of the total number of droplets per breaking event. In Part 1, it was shown that the two droplet producing sub regions within Region I-B have large standard deviations of the surface shape while the standard deviations of surface shape in Region I-A are relatively small. The remainder of droplets are measured in Region II and are generated by small bubbles that rise to the surface and burst in the wake of the breaking crest.

The distributions of droplet diameter and 2D velocity are also presented and discussed. The numbers of droplets in all diameter ranges generally increase with breaker intensity. Power law functions are fitted separately to the small and large diameter regions of these distributions and the intersections of the resulting two straight lines on the log-log plots was found to increase monotonically from 820 µm for the weak breaker to 1480 µm for the strong breaker, while the power-law exponents are approximately 2.1 and 5.8 for the large and small droplet diameter regions, respectively. This two-power law distribution shape is most strongly seen in the distributions for the moderate and strong breakers. Analysis of the droplet velocity data indicates that the average speeds are less that 1.4 m/s while the plunging jet tip impact speeds are, as given in Part 1, approximately 2.0 m/s for the three breakers. The droplet velocity directions are on average nearly vertical in Region I-A, 34° downstream from vertical in Region I-B and 16° downstream from vertical in Region II. It is speculated that the downstream components of the average velocities in Regions I-B and II are associated with the local surface drift layer created by breaking and the wave induced orbital particle motions in the breaking and following crests.

The measured droplet diameter, velocity, position and time data was used with the ensemble average profile history (form Part 1) and a low-order particle trajectory model to predict the locations where and when the droplets were generated at the free surface and the time of flight to the measurement plane. The model predicted that a good portion of the droplets measured over Region I-B were generated on the smooth water surface downstream (in the direction of wave travel) of the breaking front. It is believed that this unrealistic result is due to omission of the breaker-induced air velocity field in the droplet motion model.

**Acknowledgements.** The authors thank Prof. Joseph Katz for providing his GPU-based hologram reconstruction algorithm. The authors thank undergraduate students Benjamin Schaefer, Benjamin Cha and Nicholas Lawson for help with the image processing related to these experiments.

**Funding.** The support of the Division of Ocean Sciences of the National Science Foundation under Grants OCE0751853, OCE1829943 and OCE1925060 is gratefully acknowledged.

**Declaration of interests.** The authors report no conflict of interest.